\documentclass[11pt]{iopart}
\usepackage{graphicx,color,soul,amssymb}
\usepackage[switch]{lineno}

\begin{document}
%\begin{linenumbers}
\title{Role of work in matter exchange between finite quantum systems}

\author{Euijin Jeon$^1$, Peter Talkner$^2$, Juyeon Yi$^3$ and Yong Woon Kim$^1$\footnote{corresponding author: y.w.kim@kaist.ac.kr}}
\address{$^1$ Graduate School of Nanoscience and Technology, Korea Advanced Institute of Science and Technology, Deajeon 34141, Korea}
\address{$^2$ Institut f\"ur Physik, Universit\"at Augsburg, Universit\"atsstra{\ss}e 1, 86159 Augsburg, Germany}
\address{$^3$ Department of Physics, Pusan National University, Busan 46241, Korea}
%\ead{peter.talkner@physik.uni-augsburg.de}
%\ead{y.w.kim@kaist.ac.kr}

\date{\today}

\begin{abstract}
Close to equilibrium, the exchange of particles and heat between macroscopic systems at different temperatures and different chemical potentials is known to be governed by a matrix of transport coefficients which is positive and symmetric. We investigate the amounts of heat and particles that are exchanged between two small quantum systems within a given time, and find them characterized by a transport matrix which neither needs to be symmetric nor positive. At larger times  even spontaneous transport can be observed in the total absence of temperature and chemical potential differences provided that the two systems are different in size. All these deviations from standard transport behavior can be attributed to the fact that work is done on the system in the processes contacting and separating those parts of the system that initially possess different temperatures and chemical potentials. The standard transport properties are recovered for vanishing work and also in the limit of large systems and sufficiently large contact times. The general results are illustrated by an example. 
\end{abstract}

%\keywords{Onsager relations, transports}
%\maketitle

\section{Introduction}
The exchange of quantities such as energy and particle numbers between different parts of a spatially extended system is a fundamental phenomenon of physics, chemistry and biology. 
Being characteristic for systems out of equilibrium an exchange is typically driven by a bias of affinities such as temperature or chemical potential differences and manifests itself in the form of heat or particle currents~\cite{ashcroft,deGrootMazur}. The traditional treatment of transport phenomena is based on the notion of local equilibrium. It is formulated in terms of transport equations relating thermodynamic forces, which are caused by affinity biases, to fluxes which are defined as the time rates of change of 
the average exchanged heat and particle number.

The recent characterization of transport phenomena in terms of fluctuation relations 
\cite{bochkov,jarzynski2004,saito2008,andrieux2009,Esposito2009,Campisi2010,Campisi2011} provides an alternative understanding from a more statistical mechanical and less phenomenological point of view.
For the sake of simplicity we restrict ourselves to energy and particle exchange between two systems, A and B, each of which is initially isolated and prepared in grand-canonical equilibrium states with temperatures $\beta_A, \beta_B$ and chemical potentials $\mu_A,\mu_B$. 
If the systems are brought into contact for a certain amount of time 
during which energy and particles can be interchanged,  
the joint probability density function (pdf) $P_{\Delta E}(\Delta E_A, \Delta E_B, \Delta N)$ of the energy and particle number changes $\Delta E_\alpha$ and $\Delta N_\alpha$, of both systems, $\alpha= A,B$, respectively, obeys the following exact symmetry relation: 
\begin{equation}\label{tft}
\frac{P_{\Delta E}(\Delta E_{A},\Delta E_{B}, \Delta N)}{P_{\Delta E}(-\Delta E_{A}, -\Delta E_{B}, -\Delta N)}=\prod_{\alpha=A,B}
e^{\beta_{\alpha}(\Delta E_{\alpha}-\mu_{\alpha}\Delta N_{\alpha})}~.\end{equation}
In deriving the above relation, it is assumed that the total number of particles is conserved, and hence $\Delta N_{A}=-\Delta N_{B}\equiv \Delta N$. 
On the other hand, there is no restriction on the total energy change; the energy of the whole system may not be conserved because the contact is mediated by switching an interaction Hamiltonian on and off. In this way, work is done on both systems when they are being connected and separated. 
Thus, $\Delta E_{A}$ and $\Delta E_{B}$ are left as separate, independent variables.

The work $W$ supplied to the total system and the heat $Q$ being transferred from B to A can be expressed in terms of linear combinations  of energy and particle number differences $E_\alpha$ and $\Delta N$ \cite{jarzynski2004} as 
\begin{equation}\label{wandq}
W=\Delta E_{A}+\Delta E_{B},~Q=(\Delta E_{A}-\Delta E_{B})/2-{\bar \mu} \Delta N,
\end{equation}
where ${\overline \mu}=(\mu_{A}+\mu_{B})/2$. Equation (\ref{tft}) can be rewritten for the joint probability $P_{WQ}(W,Q,\Delta N)$ of $Q$, $W$ and $\Delta N$ yielding 
\begin{equation}\label{tft2}
P_{WQ}(W,Q, \Delta N)= e^{\overline{\beta} W + \Delta \beta Q -\overline{\beta} \Delta \mu \Delta N}P_{WQ}(-W,-Q,-\Delta N) 
\end{equation}
with ${\overline {\beta}}=(\beta_{A}+\beta_{B})/2$, $\Delta \beta \equiv \beta_{A}-\beta_{B}$ and $\Delta \mu \equiv \mu_{A}-\mu_{B}$. For an alternative definition of the affinity biases see \cite{alternative}. 
The corresponding joint pdf $P_{WQ}(W,Q,\Delta N)$ can be expressed in terms of $P_{\Delta E}(\Delta E_A, \Delta E_B, \Delta N)$ as
\begin{equation}
P_{WQ}(W,Q,\Delta N) = P_{\Delta E}(\frac{1}{2}W+Q+\bar{\mu} \Delta N, \frac{1}{2} W - Q -\bar{\mu}\Delta N, \Delta N)\:.
\label{PQ}
\end{equation}
The fluctuation relation (\ref{tft2}) was derived by various authors~\cite{andrieux2009,Esposito2009,Campisi2010}.
In most of these works, see in particular~\cite{andrieux2009},  it is assumed that (i) the two systems are large and (ii) the time $\tau$ during which the  interaction between the two systems is effective is sufficiently large so  that a quasi-stationary state with constant fluxes will prevail during most of the time. In other words, the heat as well as the exchanged particle number become proportional to the interaction-time, whereas the work, which is determined by the short periods when the interaction is turned on and off again, is independent of $\tau$. Under these conditions the work can be neglected in comparison to the heat. As a consequence, for the averages of heat $\langle Q \rangle$ and exchanged particle numbers $\langle \Delta N \rangle$, one then finds the standard directionalities $\Delta \beta \langle Q \rangle\geq 0$ if $\Delta \mu = 0$ and $-\langle \Delta N \rangle \Delta \mu \geq 0$ if $\Delta \beta =0$. For small biases the averages of the transported quantities become  linear in the affinities. i.e. $\langle Q \rangle = K_{11} \Delta \beta + K_{12} (-\bar{\beta} \Delta \mu)$ and 
$\langle \Delta N \rangle = K_{21} \Delta \beta + K_{22} (-\bar{\beta} \Delta \mu)$ hold. 
These equations are akin to standard transport equations \cite{deGrootMazur} with the difference that they describe finite amounts of exchanged heat and particle number rather than the respective fluxes. However, in the limit of large $\tau$, the transport matrix $(K_{ij})$ corresponds to  
$\tau$ times Onsager's transport matrix that relates the forces to the fluxes. 
The  symmetry and positivity of the transport matrix $(K_{i,j})$, which will be reviewed below,  
therefore imply the corresponding properties of the Onsager matrix, entailing both the reciprocity relation of Onsager's transport coefficients, in short known as reciprocity relation, and the directionality of the fluxes, which follows from the positivity of the Onsager matrix.

One though has to keep in mind that, in particular for small systems, the establishment of a long-lived quasi-stationary state may not be achieved at all or does not prevail long enough. Therefore it is not always justified to neglect the work in comparison to the heat. We shall discuss that both the directionality of the heat and particle flux, i.e. the positivity of the matrix of transport coefficients, as well as Onsager's reciprocity relations may then be violated.

The validity of Onsager's reciprocity relations is a subject that has been repeatedly discussed in the literature \cite{meixner}. The standard justification \cite{Onsager,deGrootMazur} is based on a combination of  
microscopic and phenomenological arguments \cite{Tisza,Zwanzig,Grmela}:
Microscopic reversibility is one pillar to which comes as the second pillar the assumption of a Gaussian and Markovian dynamics of the considered set of variables such as heat and particle number in the present situation. 
The Gaussian property can be justified by the assumption that only processes close to thermal equilibrium are considered. The Markovian dynamics is imposed by Onsager's regression hypothesis \cite{Onsager} which postulates that the same dynamical laws are governing  the mean values and the spontaneous fluctuations. The  relevance of the Markovian assumption  for the validity of the reciprocal relations was illustrated in Ref.~\cite{Titulaer} by an example. 

The more recent characterization of transport phenomena by means of the fluctuation theorem also requires microscopic reversibility but does not make use of the assumptions regarding the dynamics of heat and particle number: There is no restriction requiring that the considered transport processes should take place close to thermal equilibrium and hence had to stay in the linear regime; nor is there any requirement restricting the dynamics of the transported quantities apart from microscopic reversibility of the underlying microscopic dynamics as already emphasized. 
The second essential postulate  assumes the existence of a long-lived quasi-stationary current carrying state requiring sufficiently large systems and a large contact time during which the two systems quickly build up 
the mentioned quasi-stationary state.
    
The paper is organized as follows: In Sec. II, the general setting is specified: Two initially isolated systems each of which is in thermal equilibrium with possibly different temperatures and chemical potentials are brought in contact and again separated after some time $\tau$. For this setting Eq.~(\ref{tft}) is obtained. With work and heat identified as in Eq. (\ref{wandq}) the fluctuation theorem (\ref{tft2}) is expressed in terms of these quantities.        
Section III is devoted to the
discussion of the implications of work on the exchange properties. In a first subsection, we consider the case of vanishing work, $W=0$, and 
demonstrate how the directionality and the Onsager relations follow from Eq.~(\ref{tft2}).
For non-vanishing work we find that the transport matrix is asymmetric and need also not be positive. Moreover, a spontaneous flow of heat and particles may occur even in the absence of any affinity bias  as presented in the second subsection. 
In Sec. IV, we consider an example which illustrates the findings of Sec. III. A Summary concludes the paper in Sec. V. 

\section{Setting}
We consider a total system that consists of two subsystems $A$ and $B$ with Hamiltonians $\hat{H}_A$, $\hat{H}_B$ and particle number-operators $\hat{N}_A$, $\hat{N}_B $, respectively. For the sake of simplicity we consider only one sort of particles that may reside in both systems.
Up to the time $t=0$ the systems $A$ and $B$ are isolated from each other and stay in states of grand-canonical equilibrium. The initial density matrix is therefore given by 
\begin{equation}\label{rho}
\hat{\rho}(0)=\prod_{\alpha=A,B}e^{-\beta_\alpha(\hat{H}_{\alpha}-\mu_{\alpha}\hat{N}_{\alpha})}/\mathcal{Z}_\alpha
\end{equation}
with $\mathcal{Z}_\alpha$ being the grand-canonical partition function of system $\alpha = A,B$.
The affinity parameters, i.e. the inverse temperature $\beta_\alpha$ and the chemical potential $\mu_\alpha$, can in general be  different: $\beta_A\neq \beta_B$ and $\mu_A\neq \mu_B$.
The Hamiltonians $\hat{H}_\alpha$ and particle number operators $\hat{N}_\alpha$ commute with each other, and, consequently 
have simultaneous eigenstates with the corresponding eigenvalues $E_\alpha$ and $N_\alpha$ satisfying: $\hat{H}_{\alpha}|E^\alpha_i,N^\alpha_i\rangle=E^{\alpha}_i|E^\alpha_i,N^\alpha_i\rangle$ and $\hat{N}_{\alpha}|E^\alpha_i,N^\alpha_i\rangle=N^{\alpha}_i|E^\alpha_i,N^\alpha_i\rangle$. For the sake of simplicity we assume that these states are non-degenerate. 
A basis in the Hilbert space of the total system $A\oplus B$ is spanned by the product states  
$|E^A_i,N^{A}_i\rangle \otimes |E^B_i,N^B_i\rangle \equiv |i\rangle$, where the index $i$ stands for a complete set of quantum numbers. 
The probability $P_i$ to find the set of quantum numbers $E^A_i,E^B_i,N^A_i,N^B_i$ 
in a joint measurement of $\hat{H}_A,\;\hat{H}_B,\;\hat{N}_A,\;\hat{N}_B$ in the initial state $\hat{\rho}(0)$ 
is given by $P_i= \langle i |\hat{\rho}(0)|i\rangle $ and hence becomes
\begin{equation}\label{inieq}
P_{i}=\prod_{\alpha=A,B}e^{-\beta_{\alpha}(E^{\alpha}_i-\mu_{\alpha}N^{\alpha}_i)}/{\cal Z}_{\alpha}~.
\end{equation}

We suppose that a coupling between the two systems, described by the interaction Hamiltonian $\hat{H}_{C}$, is turned on at time $t=0^{+}$.  The quantum state of the total system $A\oplus B$ subsequently evolves in time (for $t>0$) according to the total, time-reversal invariant Hamiltonian \cite{timereversal} 
\begin{equation}
\hat{H}=\hat{H}_{A}+\hat{H}_{B}+\hat{H}_{C}\:.
\label{Htot}
\end{equation}
When the time $t=\tau^{-}$ has elapsed the coupling is switched off and immediately after,  the energy and the particle number of each system are measured. This measurement projects the system state onto a common eigenstate of $\hat{H}_\alpha$ and $\hat{N}_\alpha$,  $|f\rangle\equiv|E^A_f,N^A_f\rangle \otimes|E^B_f,N^B_f\rangle$, where  $E^{\alpha}_f$ and $N^{\alpha}_f$ are 
eigenvalues of the Hamiltonians and particle number operators of the respective isolated systems.

The exchange of energy and particles between the systems can be quantified in terms of the measured eigenvalues by $\Delta E^\alpha=E^{\alpha}_f-E^{\alpha}_i$ and $\Delta N=N^{A}_f-N^{A}_i=-N^{B}_f+N^{B}_i$, respectively, with $E^\alpha_j$ and $N^\alpha_j$ being eigenvalues of the Hamiltonians and particle number operators obtained in the initial ($j=i$) and the final ($j=f$) measurements.
The joint probability to find certain values of $\Delta E_\alpha$ and $\Delta N$ is determined by 
\begin{equation}
P_{\Delta E}(\Delta E_A, \Delta E_B, \Delta N)= \langle  \delta(\Delta N - N^A_f +N_i^A) \prod_{\alpha = A,B} \delta(\Delta E_\alpha- E^\alpha_f + E^\alpha_i) \rangle\:,
\label{pdf1}
\end{equation}
where $\langle \cdot \rangle = \sum_{i,f} \cdot p(f,i)$ denotes the average with respect to the joint probability $p(f,i)$ to find the set of eigenvalues $i$ in the first and the set $f$ in the second measurement. 
The sum runs over all sets of eigenvalues $i$ and $f$ and $p(f,i)$ is given by
\begin{equation}
p(f,i) = T_{f|i} P_i\:,
\label{pfi}
\end{equation}
where $P_i$ is defined in Eq. (\ref{inieq}) and $T_{f|i}$ denotes the transition probability between the states $|i \rangle $ and $|f \rangle$ and, hence, is given by
\begin{equation}
	T_{f|i}=|\langle  f|\hat{U}(\tau)|i\rangle|^2.
\label{Tfi}
\end{equation}
The unitary time-evolution $\hat{U}(\tau) = e^{-i \hat{H} \tau/\hbar}$ propagates the state of the total system  
from $t=0$ until $t=\tau$ in terms of the total Hamiltonian $\hat{H}$ specified by Eq. (\ref{Htot}).
We remark that the probabilities $P_i$ and $P_f$ of any two energy and particle number configurations $i$ and $f$ are 
related to each other by 
\begin{equation}\label{finaleq}
P_{i}= e^{M_{f,i}}P_{f} \:,
\end{equation} 
where 
\begin{equation}
M_{f,i}\equiv \sum_{\alpha}\beta_{\alpha}\left [(E^{\alpha}_f-E^\alpha_i) -\mu_{\alpha} (N^{\alpha}_f- N^\alpha_i) \right]
\label{Mif}
\end{equation} 
depends only on the energy differences between the two configurations and the according number difference. Note that there is no summation on the index $f$ in Eq. (\ref{finaleq}). Note further that the exponential function of $M$ appears on the right hand side of the fluctuation relation (\ref{tft}). Owing to the time-reversal invariance of the Hamiltonian, 
the transition probability is symmetric with respect to the initial and the final state:
\begin{equation}\label{mtr}
T_{f|i}=|\langle  f|\hat{U}(\tau)|i\rangle|^2=|\langle  i|\hat{U}(\tau)|f\rangle|^2=T_{i|f}~.
\end{equation}
Similar detailed balance like relations can also be derived under more general conditions \cite{talkner2012}. As long as the protocol specifying how the interaction between the systems $A$ and $B$ is switched on and off is symmetric in time and the indices $i$ and $f$ specify non-degenerate states the simple form (\ref{mtr}) holds unchanged.   

Together with Eq. (\ref{mtr}) the Eq. (\ref{finaleq}) implies an analogous relation for the joint probability $p(i,j)$ reading
\begin{equation}
p(i,f) = e^{M_{i,f}} p(f,i)\:.
\label{pMp}
\end{equation} 
Combined with Eq. (\ref{pdf1}) it yields
{\setlength{\mathindent}{0cm}
\begin{eqnarray}\nonumber
P_{\Delta E}(\Delta E_A, \Delta E_B, \Delta N) &=& e^{\sum_\alpha \beta_\alpha(\Delta E_\alpha -\mu_\alpha \Delta N_\alpha)}\sum_{i,f} T_{i|f} P_{f} \delta ( \Delta N - N^A_{f} +N^A_{i}) \\ \nonumber
&&\times\prod_{\alpha=A,B}\delta(\Delta E_{\alpha}-E^{\alpha}_f +E^{\alpha}_i) \\ 
&=&e^{\sum_\alpha \beta_\alpha(\Delta E_\alpha -\mu_\alpha \Delta N_\alpha)} 
P_{\Delta E}(-\Delta E_A\, -\Delta E_B, -\Delta N)~\label{tftprove},
\end{eqnarray}}
where the last equality is obtained by interchanging $i$ and $f$. 
The above equation proves the fluctuation theorem (\ref{tft}), which can be transformed into Eq.~(\ref{tft2}) upon using the definitions of work and heat, 
presented in Eq.~(\ref{wandq}). It is worth noticing that, due to the symmetry of the switching process, the probabilities on both sides of Eq. (\ref{tftprove}) and consequently also those entering the fluctuation relation (\ref{tft}) refer to the same process.

\section{Generalities}
Next we extract the essential properties of heat and number exchange 
from the particular form of the joint particle and energy probability density (\ref{pdf1}) and the fluctuation theorem~(\ref{tft2}). 
In order to better understand the role played by  the work, we first assume a situation in which  
the work vanishes. In this particular case, the expected directionality of heat from hot to cold and of the particle flux from high to low chemical potential follows. For small affinity differences, the exchanged heat and particle number are linearly related to the affinities with coefficients satisfying Onsager's symmetry relation. In cases in which the work is finite and cannot be neglected in comparison to the exchanged heat or the energy related to particle transport, both properties need not hold any longer.

\subsection{Energy conserving process: $W=0$}
We here consider processes for which the work done by turning the interaction on and off 
can be neglected compared to the exchanged heat and the energy transported by the exchanged particles. 
The joint work-heat-number pdf can then be approximated by 
$P_{WQ}(W,Q,\Delta N)\approx \delta(W)P_{Q}(Q,\Delta N)$, 
where $P_{Q}(Q,\Delta N)$ satisfies a reduced form of the fluctuation theorem, 
\begin{equation}\label{W0FT}
	P_Q(Q,\Delta N)=e^{\Delta \beta Q-\bar{\beta}\Delta\mu\Delta N}P_Q(-Q,-\Delta N)\:.
\end{equation}
The integration of both sides of Eq.(\ref{W0FT}) over $Q$ and $\Delta N$ yields the  identity 
$\langle e^{-\Delta\beta Q+\bar{\beta}\Delta\mu \Delta N}\rangle=1$.  With Jensen's inequality, $\langle e^{x}\rangle\ge e^{\langle x\rangle}$, one obtains
\begin{equation}\label{dir}
	\Delta \beta\langle Q\rangle-\bar{\beta}\Delta\mu \langle\Delta N\rangle\ge 0 \:.
\end{equation}
This implies the directionality of matter exchange: The average of heat, $\langle Q\rangle$, induced by a positive $\Delta\beta$ with $\Delta\mu=0$, is nonnegative,
indicating that the part with higher temperature looses energy which is transferred to the part with lower temperature, as stated below Eq.~(\ref{tft2}).
Also, in the absence of a temperature difference ($\Delta \beta=0$), from Eq.~(\ref{dir})  
the average number change caused by a finite chemical potential difference 
satisfies $\Delta \mu \langle \Delta N\rangle \leq 0$. This indicates that particles are transported from the region 
initially at high chemical potential to the region at lower chemical potential. Note that the directions into which heat and particles are transported only depend on those of the affinity biases but not on their sizes provided the work $W$ can be neglected.     

For processes driven by small affinities one may expand the exponential factor on the right hand side of Eq. (\ref{W0FT}) to yield
\begin{equation}
P_Q(Q,\Delta N) = (1 + \Delta \beta \:Q  - \bar{\beta} \Delta \mu \:\Delta N) P_Q(-Q,-\Delta N)\:.
\label{W0lr}
\end{equation}
Multiplying both sides by either $Q$ or $\Delta N$ and integrating over $Q$ and $\Delta N$ one obtains  
\begin{eqnarray}
	\langle Q\rangle&=&\frac{1}{2}[\langle Q^2\rangle_{Q0} \Delta \beta -\langle Q\Delta N\rangle_{Q0}\bar{\beta}\Delta \mu] \label{W0OnsagerQ}\\
	\langle \Delta N\rangle &=&\frac{1}{2}[\langle Q\Delta N\rangle_{Q0} \Delta \beta -\langle \Delta N^2\rangle_{Q0} \bar{\beta}\Delta \mu]\:,\label{W0OnsagerN}
\end{eqnarray}
where $\langle \cdot \rangle_{Q0} = \int dW d\Delta N \cdot P_Q(Q,\Delta N)$ represents the average in the absence of any bias ($\Delta\mu=\Delta\beta=0$). 

These equations are akin to standard transport equations with the difference that the latter describe the transport behavior in terms of instantaneous fluxes caused by the momentary affinity biases while the Eqs. (\ref{W0OnsagerQ}) and (\ref{W0OnsagerN}) quantify the total amounts of exchanged heat and particle numbers. In the standard transport equations, the time $t$ may take any value from the beginning $t=0$ until the end $t=t_f$ of the considered experiment, in contrast to the above relations, in which $\tau=t_f$ only refers to the immediate instant of time after which the interaction is turned off. For transport phenomena in quantum systems, a continuous observation of the fluxes is not feasible because the unavoidable back-action of the necessary measurement would have a too strong impact on the result. A similar situation is met with work measurements as discussed in Ref.~\cite{Venkatesh}. Also there, the least invasive process diagnosis is given by two energy measurements, one immediately before and the second one immediately after the process is completed.     

With the assumption that after a time $t\ll \tau$ a quasi-stationary state has established \cite{andrieux2009,Campisi2010}, the heat and particle number fluxes, $\dot{Q}$ and $\Delta \dot{N}$,  can be inferred from the totally exchanged heat and particle number as $\dot{Q} = Q/\tau$ and $\Delta\dot{N} =\Delta N/\tau$, respectively. Accordingly, with the definition of transport coefficients $L_{11} = \langle Q^2 \rangle_{Q0}/\tau,\; L_{12}=L_{21} = \langle Q \Delta N \rangle_{Q0}/\tau,\; L_{22} = \langle (\Delta N)^2 \rangle_{Q0}/\tau$ one recovers from the Eqs. (\ref{W0OnsagerQ}) and (\ref{W0OnsagerN}) the standard form of linear  transport equations with a symmetric matrix of transport coefficients in accordance with Onsager's symmetry  relations. The positivity of the matrix follows immediately from the fact that it is proportional to the covariance matrix of $Q$ and $\Delta N$ with the positive proportionality factor $1/\tau$.      

In summary, we find that, with the Eqs. (\ref{W0OnsagerQ}) and (\ref{W0OnsagerN}), the total amounts of heat and exchanged particles follow the standard transport rules concerning symmetry and directionality provided that the work applied to the system vanishes or can be neglected. 

\subsection{Energy non-conserving processes: $W\neq 0$}\label{n0W}

Now we focus our considerations to processes
in which work is performed on the total system, and hence its energy differs at the end from what it was in the beginning.

We will still assume that the affinity biases $\Delta \beta =\beta_A-\beta_B$ and $\Delta \mu = \mu_A- \mu_B$ are small compared to their average values $\bar{\beta} = (\beta_A + \beta_B)/2$ and $\bar{\mu} = (\mu_A + \mu_B)/2$, respectively. The only dependence of the joint probability density function $P_{WQ}(W,Q,\Delta N)$ on the affinity biases is contained in the initial distribution of energies and particle numbers, $P_i$ given by Eq. (\ref{inieq}). For small affinity biases $P_i$ can be expanded yielding up to first order in $\Delta \beta$ and $\bar{\beta} \Delta \mu$ the following expression
\begin{equation}
P_i =\left ( 1 - X^\beta_i \Delta \beta - X^\mu_i (-\bar{\beta} \Delta \mu ) \right ) P^0_i\:,
\label{Pi}
\end{equation}
where $P^0_i$ is the probability to find the systems $A$ and $B$ at equal temperature $\bar{\beta}$ and chemical potential $\bar{\mu}$, hence reading
\begin{equation}
%P^0_i = \prod_{\alpha=A,B} e^{-\bar{\beta} (E^\alpha_i -\bar{\mu} N^\alpha_i)} \mathcal{Z}^{-1}_\alpha\:.
P^0_i = \frac{1}{{\mathcal Z}^{0}}\prod_{\alpha=A,B}e^{-\bar{\beta} (E^\alpha_i -\bar{\mu} N^\alpha_i)}
\label{Pi0}
\end{equation}
with ${\mathcal Z}^{0}$ being the corresponding grand canonical partition function.
The coefficients in front of the affinity biases are given by
\begin{eqnarray}
\label{Xb}
X^\beta_i &=& \frac{1}{2} \left ( \delta E^A_i -  \delta E^B_i - \bar{\mu} \delta N^A_i + \bar{\mu} \delta N^B_{i} \right )\\
X^\mu_i &=& \frac{1}{2} \left ( \delta N^A_i - \delta N^B_{i} \right )\:.
\label{Nm}
\end{eqnarray}
Here $\delta E^\alpha_i = E^\alpha_i - \overline{E^\alpha}$ and $\delta N^\alpha_i = N^\alpha_i - \overline{N^\alpha}$ denote fluctuations of energies and particle numbers about their averages in the bias-free initial state given by $\overline{E^\alpha} = \sum_i E^\alpha_i P^0_i$ and  $\overline{N^\alpha} = \sum_i N^\alpha_i P^0_i$, respectively. 

The fluctuating heat and particle number exchange can be expressed as differences between the values of $X^\beta_j$ and $X^\mu_j$ as they result from the first ($j=i$) and the final ($j=f$) energy and particle measurements, yielding
\begin{eqnarray}
\label{QX}
Q &=& X^\beta_f - X^\beta_i\\
\Delta N &=& X^\mu_f - X^\mu_i\:.
\label{NX}
\end{eqnarray}
Accordingly, the averages of $Q$ and $\Delta N$  become
\begin{eqnarray}
\label{QXp}
\langle Q \rangle &=& \langle X^\beta(\tau) -X^\beta(0)\rangle \\
&=& \sum_{i,f} (X^\beta_f - X^\beta_i) T_{f|i} P_i \nonumber \\
\label{NQp}
\langle \Delta N \rangle &=& \langle X^\mu(\tau) -X^\mu(0) \rangle\\
&=& \sum_{i,f} (X^\mu_f - X^\mu_i) T_{f|i} P_i \:. \nonumber 
\end{eqnarray}
Replacing now the initial probability $P_i$ by its small affinity bias approximation (\ref{Pi}) one obtains for these averages up to first order in $\Delta \beta$ and $\Delta \mu$ 
\begin{eqnarray}
\label{Q1}
\langle Q \rangle &=& \langle X^\beta(\tau)-X^\beta(0) \rangle_0 + C_{\beta,\beta} \Delta \beta +C_{\beta,\mu} (-\bar{\beta} \Delta \mu)\\
\label{N1}
\langle \Delta N \rangle  &=& \langle X^\mu(\tau)-X^\mu(0) \rangle_0 + C_{\mu,\beta} \Delta \beta +C_{\mu,\mu} (-\bar{\beta} \Delta \mu)\:,
\end{eqnarray}
where $\langle \cdot \rangle_0 = \sum_{i,f} \cdot T_{f|i} P^0_i$ and 
\begin{equation}
C_{\chi,\eta} = - \langle \left [X^\chi(\tau) - X^\chi(0) \right ] X^\eta(0) \rangle_0\:.
\label{C}
\end{equation}
This result differs in two respects from Onsager's standard transport theory:
First, heat and particles may be exchanged between the two systems even if the affinity biases vanish, and second, the matrix $C = (C_{\chi,\eta})$ governing the transport caused by small affinity differences needs neither be symmetric nor positive.

As already demonstrated above, in the absence of an affinity bias transport does not occur if the energy of the total system remains constant under the influence of the coupling between $A$ and $B$. 
This can also be seen from the symmetry relation (\ref{pMp}) of the joint probability which for vanishing biases $\Delta \beta =0$ and $\Delta \mu =0$
simplifies to
\begin{equation}
p^{\Delta \beta =0,\Delta \mu =0}(i,f) = e^{ \bar{\beta} W_{f,i}} p^{\Delta \beta =0,\Delta \mu =0}(f,i)\:,
\label{pWp}
\end{equation}
where $W_{f,i} = \sum_\alpha E^{\alpha}_f - E^{\alpha}_i$ is the work performed on the system upon a transition from $i$ to $f$.
When this work vanishes for all possible transitions, i.e. for all those pairs $i,f$ with $p(f,i) \ne 0$,  the transition probability is symmetric. Then, the averages of $X^\chi(t)$ agree at $t=0$ and $t=\tau$, and hence we recover that any transport of heat and particles may only occur due to affinity biases but not spontaneously. Here, the index $\chi$ may refer to $\beta$ or $\mu$. 

Because in the presence of finite work, the joint probability is no longer stationary, the averages $X^\chi(\tau)$ and $X^\chi(0)$ will in general disagree, as already noted. Moreover, due to the non-stationarity of the joint probability $p(i,f)$ the auto-correlation functions $\langle X(\tau) X(0)\rangle_0$ may become larger than the second moment $\langle X^2(0) \rangle_0$, where $X=a X^\beta +b X^\mu$, $a,b$ real, is an arbitrary linear combination of $X^\beta$ and $X^\mu$. With $\langle X^2(0) \rangle_0 \leq \langle X(\tau)X(0) \rangle_0$ the matrix $C =(C_{\chi,\eta})$ is no longer positive and hence the directionality of the affinity bias induced transport may also differ from the standard Onsager rules.

Moreover, the reciprocity relations are in general violated by the presence of work rendering the matrix $C$ non-symmetric because in general $\langle X^\beta(\tau) X^\mu(0) \rangle_0 \ne  \langle X^\mu(\tau) X^\beta(0) \rangle_0$ and hence $C_{\beta,\mu} \ne C_{\mu,\beta}$.

A positive and symmetric matrix $L$ determines the averages of  heat and exchanged particle numbers if they are modified by the factor $(1+e^{-\bar{\beta} W})/2$. Multiplying both sides of the fluctuation relation (\ref{tft2}) by $e^{-\bar{\beta}W}$ and optionally by either $Q$ or $\Delta N$ and integrating over all $W$, $Q$ and $\Delta N$ one obtains to first order in the affinity biases the expressions
\begin{eqnarray}
%\begin{split}
\langle Q (e^{-\bar{\beta} W} +1) \rangle &=& 2 \left ( L_{\beta,\beta} \Delta \beta + L_{\beta,\mu} (-\bar{\beta} \Delta \mu) \right ) \nonumber \\  
\langle \Delta N (e^{-\bar{\beta} W} +1) \rangle &=& 2 \left ( L_{\mu,\beta} \Delta \beta + L_{\mu,\mu} (-\bar{\beta} \Delta \mu) \right )\:,
%\end{split}
\label{QL}
\end{eqnarray}
where 
\begin{equation}
2 L_{\chi,\eta} = \langle (X^\chi(\tau) -X^\chi(0))(X^\eta(\tau) -X^\eta(0)) \rangle_{0}
\label{LXX}
\end{equation}
coincides with the covariance matrix of $Q$ and $\Delta N$ when there is no affinity bias. 
The deviation $D$ of the actual transport matrix $C$ from $L$, $D=L-C$, is given by 
\begin{equation}
2 D_{\chi,\eta} =  \langle (X^\chi(\tau) -X^\chi(0))( X^\eta(\tau) + X^\eta(0) ) \rangle_0\:.
\label{D}
\end{equation} 

We conclude that the work that is performed when the two parts of the system are brought in contact and finally are disconnected again gives rise to several unexpected effects such as spontaneous transport, non-reciprocal cross-terms of the transport matrix and deviations from the standard directionalities of transport. The work which causes these anomalies is a measure for the amount of the non-stationarity imposed by the time limitation of the transport experiment.

In the following section, we consider a particular example and demonstrate that these deviations from the conventional transport theory 
appear also for weakly coupled systems provided they are not too large. 

\subsection{Model system}

In order to substantiate the existence of the nontrivial effects of work we consider as parts $A$ and $B$ two Fermionic systems that are described by tight-binding Hamiltonians of the form:
\begin{equation}
\hat{H}_\alpha=-\gamma\sum_{x_\alpha=1}^{M_\alpha-2}[{c}_{x_\alpha}^\dagger c_{x_\alpha+1}
+c^{\dagger}_{x_\alpha+1}c_{x_\alpha}]\:,\quad \alpha = A,B\:,
\label{Ha}
\end{equation} 
where $M_\alpha -1$ denotes the number of sites of the part $\alpha$ and the operator $\hat{c}_{x_\alpha}$ ($\hat{c}_{x_\alpha}^{\dagger}$) annihilates (creates) a fermion at the site $x_{\alpha}$ of the system $\alpha$.
The hopping energy $\gamma$ determines the energy scale of the system. We assume that the partial systems A and B are identical except that they may have different $M_\alpha$.
The particle numbers in each system are specified by operators $\hat{N}_{\alpha}$ defined by
\begin{equation}
\hat{N}_\alpha=\sum_{x_\alpha=1_\alpha}^{M_\alpha-1}\hat{c}^{\dagger}_{x_\alpha}\hat{c}_{x_\alpha}.
\end{equation}
The coupling Hamiltonian which is turned on at $t=0^+$ is given by 
%\begin{equation}
%\hat{H}_C=-\gamma_C \hat{c}_{1_A}^\dagger \hat{c}_{1_B}-\gamma_C \hat{c}_{1_B}^\dagger \hat{c}_{1_A}\:.
%\end{equation}
\begin{equation}
\hat{H}_C=-\gamma_C \left (\hat{c}_{1_A}^\dagger \hat{c}_{1_B} + \hat{c}_{1_B}^\dagger \hat{c}_{1_A} \right) \:.
\end{equation}
It connects the two end sites $1_A$ and $1_B$ enabling the exchange of  particles between the two systems under the constraint of a constant total particle number.

These Hamiltonians describe various physical systems, like systems of electrons with negligible  spin-degrees of freedom~\cite{spinless},
hard-core bosons in one-dimensional optical lattices~\cite{Cazalilla2011} or quantum spin rotors~\cite{Auerbach1992}. 
This class of systems can be solved exactly because there is no interaction between particles.
We verify the presence of spontaneous flow and deviations from Onsager symmetry. 

\subsection{Method}

We are interested in the temporal changes of the energies and the  particle numbers, which are determined by the eigenvalues of the operators $\hat{H}_\alpha=\sum_{n_\alpha=1}^{M_\alpha-1}\varepsilon_{n_\alpha}{\tilde{c}}_{n_\alpha}^\dagger\tilde{c}_{n_\alpha}$ and $\hat{N}_\alpha=\sum_{n_\alpha=1}^{M_\alpha-1}{\tilde{c}}_{n_\alpha}^{\dagger}\tilde{c}_{n_\alpha}$, respectively.
Here, the Fermi operators $\tilde{c}_{n_\alpha}$ diagonalize the Hamiltonian $\hat{H}_\alpha$. They are given by
\begin{equation}
\tilde{c}_{n_\alpha}=\sum_{x_\alpha=1}^{M_\alpha-1}a_{n_\alpha,x_\alpha}\hat{c}_{x_\alpha}
\end{equation}
with the coefficients
\begin{equation}
a_{n_\alpha,x_\alpha}=\sqrt{\frac{2}{M_\alpha}}\sin\left(\frac{n_\alpha\pi x_\alpha}{M_\alpha}\right)\:.
\end{equation}
The energy eigenvalues result in $\varepsilon_{n_\alpha}=-2\gamma \cos(n_\alpha \pi/M_\alpha)$.
To obtain the time-evolution of the annihilation operators in presence of the interaction we  
consider the Heisenberg equations of motion, which are
\begin{eqnarray}
i\hbar \frac{d}{dt}\tilde{c}_{n_\alpha}(t)&=&\varepsilon_{n_\alpha}\tilde{c}_{n_\alpha}(t)+\sum_{n_{\alpha^\prime\neq\alpha}=1}^{M_{\alpha^\prime}-1}V_{n_\alpha,n_{\alpha^\prime}}\tilde{c}_{n_{\alpha^\prime}}(t)\\ \nonumber
&=&\sum_{n_{\alpha^{\prime}}=1}^{M_{\alpha^{\prime}}-1}H_{n_\alpha,n_{\alpha^{\prime}}}\tilde{c}_{n_{\alpha^\prime}}(t),
\end{eqnarray}
where $V_{n_\alpha,n_{\alpha^\prime}} =-\gamma_C a_{n_A,1_A}a_{n_B,1_B}  $ 
determines the coupling Hamiltonian written in terms of 
$\{\tilde{c}_{n_\alpha}\}$:
\begin{equation}
\hat{H}_C=\sum_{n_A,n_B}V_{n_A,n_B}(\tilde{c}_{n_A}^\dagger\tilde{c}_{n_B}+\tilde{c}_{n_B}^{\dagger}\tilde{c}_{n_A})\:.
\end{equation}
The retarded Green's function, defined by $G^r_{n_\alpha,n_{\alpha^\prime}}(t_2-t_1)=-i\langle\{\tilde{c}_{n_\alpha}(t_2),\tilde{c}_{n_{\alpha^\prime}}^\dagger(t_1)\}\rangle$ satisfies the following equation of motion:
\begin{equation}\label{Gdif}
i\hbar \frac{d}{dt} G^r_{n_\alpha,n_{\alpha^\prime}}(t)=\sum_{n_{\alpha^{\prime\prime}}=1}^{M_{\alpha^{\prime\prime}}-1}H_{n_\alpha,n_{\alpha^{\prime\prime}}}G^{r}_{n_{\alpha^{\prime\prime}}}(t)
\end{equation}
with the initial condition: $G^r_{n_\alpha,n_{\alpha^\prime}}(0)=-i\delta_{n_\alpha,n_{\alpha^\prime}}$, following from  $\tilde{c}_{n_\alpha}(t)=i\sum_{n_{\alpha^\prime}}G^r_{n_\alpha,n_{\alpha^\prime}}(t)\tilde{c}_{n_{\alpha^\prime}}(0)$. Here, $H_{n_\alpha,n_{\alpha''}}$ is the matrix element of the total Hamiltonian $\hat{H} = \hat{H}_A + \hat{H}_B + \hat{H}_C$ with respect to the site basis generated by the creation operators $c^\dagger_{x_\alpha}$.
The solution of Eq.~(\ref{Gdif}) can be evaluated by exact diagonalization of $H_{n_\alpha,n_{\alpha^{\prime\prime}}}$.
The temporal behavior of several thermodynamic quantities can be expressed utilizing the retarded Green's funtion; for example, the average amount of energy change in the system $\alpha$,
\begin{eqnarray}
\langle \Delta E_\alpha\rangle&=&\sum_{n_{\alpha}}\varepsilon_{n_\alpha}\langle \tilde{c}^{\dagger}_{n_\alpha}(\tau)\tilde{c}_{n_\alpha}(\tau)-\tilde{c}^\dagger_{n_\alpha}(0)\tilde{c}_{n_\alpha}(0)\rangle\\ \nonumber
&=&\sum_{\alpha^\prime=1}^{2}\sum_{n_\alpha,n_{\alpha^\prime}}\varepsilon_{n_\alpha} \left [|G^r_{n_\alpha,n_{\alpha^\prime}}(\tau)|^2-\delta_{n_\alpha,n_{\alpha^\prime}} \right ]f_{n_{\alpha^\prime}},
\end{eqnarray}
and the average number of particle change,
\begin{equation}
\langle \Delta N_\alpha\rangle=\sum_{\alpha=1}^{2}\sum_{n_\alpha,n_{\alpha^\prime}}\left [|G^r_{n_\alpha,n_{\alpha^\prime}}(\tau)|^2-\delta_{n_\alpha,n_{\alpha^\prime}}\right ]f_{n_{\alpha^\prime}},
\end{equation}
are written in terms of $G^r_{n_\alpha,n_{\alpha^\prime}}(\tau)$ with the Fermi-Dirac distribution function $f_{n_\alpha}=\langle \tilde{c}_{n_\alpha}^\dagger(0)\tilde{c}_{n_\alpha}(0)\rangle=[e^{-\beta_\alpha(\varepsilon_{n_\alpha}-\mu_\alpha)}+1]^{-1}$.

Then the average heat and exchanged particle number can be expressed as
\begin{eqnarray}
\label{QG}
\langle Q\rangle &=& \sum_{s,s^\prime}x^\beta_{s}\left [|G^r_{s,s^\prime}(\tau)|^2-\delta_{s,s^\prime}\right ]f_{s^\prime}\\
\label{NG}
\langle \Delta N\rangle &=&\sum_{s,s^\prime}x^\mu_{s}\left [|G^r_{s,s^\prime}(\tau)|^2-\delta_{s,s^\prime}\right ]f_{s^\prime}\:,
\end{eqnarray}
where the summation indices $s$ and $s^\prime$ run over the energy levels of both systems. 
Depending on whether $s$ indicates a level of system $A$ or $B$ the coefficients $x^\beta_s$ are defined as 
\begin{eqnarray}
%\begin{split}
x^\beta_{n_A} &=& (\varepsilon_{n_A} - \mu_A)/2 \nonumber\\
x^\beta_{n_B} &=& -(\varepsilon_{n_B} - \mu_B)/2 \:,
%\end{split}
\label{xb}
\end{eqnarray}
whereas 
\begin{equation}
x^\mu_{n_A} =1\:, \quad x^\mu_{n_B}=0 \: .
\label{xm}
\end{equation}
The covariance matrix $2L$ of $Q$ and  $\Delta N$ for $\beta_A=\beta_B=\beta$ and $\mu_A=\mu_B=\mu$ can be written as 
\begin{eqnarray}\nonumber
%\begin{split}
2L_{\chi,\eta}&=&\sum_{s,s^\prime}x^\chi_{s} x^\eta_{s^\prime}\langle \left [c_{s}^\dagger(\tau)c_{s}(\tau)-c_{s}^\dagger(0)c_s(0)\right ]\left [c^{\dagger}_{s^\prime}(\tau)c_{s^\prime}(\tau)-c_{s^\prime}^\dagger(0) c_{s^\prime}(0)\right ]\rangle_{0} \\ \nonumber
&=&\sum_{i,j,k,l,s,s^\prime}x^\chi_{s}x^\eta_{s^\prime}\left [G^{*}_{s,i}(\tau)G_{s,j}(\tau)-\delta_{s,i}\delta_{s,j}\right ] \\ 
&&\times \left [G^*_{s^\prime,k}(\tau)G_{s^\prime,l}(\tau)-\delta_{s^\prime,k}\delta_{s^\prime,l} \right ]\langle c^\dagger_{i}c_{j}c^\dagger_{k}c_{l}\rangle_{0},
%\end{split}
\label{LG}
\end{eqnarray}
where the summations extend over  all energy levels of both systems.
The equilibrium averages of the fourths moments of creation and annihilation operators can be expressed in terms of the Fermi-Dirac distribution function: $\langle c^{\dagger}_{i}c_{j}c^\dagger_{k}c_{l}\rangle_0=\delta_{i,j}\delta_{k,l}f_if_k+\delta_{i,l}\delta_{j,k}f_i(1-f_k)$,
finally yielding
\begin{eqnarray}\nonumber
2L_{\chi,\eta}&=&\langle (X^\chi(\tau)-X^\chi(0))\rangle_{0} \langle (X^\eta(\tau) -X^\eta(0)) \rangle_{0} \\ \nonumber
&+&\sum_{s,s^\prime,i,k}x^\chi_{s}x^\eta_{s^\prime}\left [G^{*}_{s,i}(\tau)G_{s,k}(\tau)-\delta_{s,i}\delta_{s,k}\right ] \\
&&\times \left [G^*_{s^\prime,k}(\tau)G_{s^\prime,i}(\tau)-\delta_{s^\prime,k}\delta_{s^\prime,i}\right ]f_i(1-f_k) \;,
\end{eqnarray}
where $\langle X^\beta(\tau) - X^\beta(0) \rangle_0 = \langle Q \rangle_0$ and  $\langle X^\mu(\tau) - X^\mu(0) \rangle_0 = \langle \Delta N \rangle_0$ according to Eqs. (\ref{QXp}) and (\ref{NQp}), respectively.

In the next Section we determine the relative deviation $D$ of the actual transport matrix $C$ from the symmetric matrix $L$.
The calculation of the second moments implies four summations. Accordingly, the time required for the calculation grows with the system size proportionally to $(M_A+M_B)^{4}$ which is still feasible  for the  relatively small systems considered here.

\subsection{Results}

We illustrate the spontaneous transport as well as the asymmetry property of the
transport matrix for relatively small systems with $M_B = 100,\;200$ and  different sizes of $A$. As in the recent study \cite{euijin2016}, the interaction $\gamma_C$ is chosen small compared to the hopping energy $\gamma$. Likewise, the chemical potentials are also relatively small compared to the hopping energy. 

\begin{figure}[tbp]
	\centering
	\includegraphics[width=\linewidth]{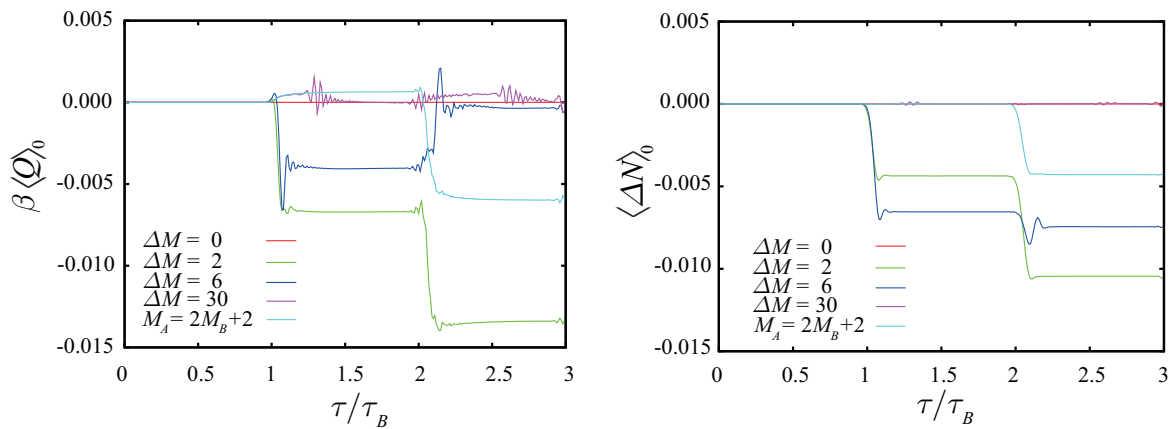}
	\caption {
The amounts of heat (left panel) and number of particles (right panel) that flow spontaneously in the absence of any affinity bias are displayed as  functions of the total time of contact $\tau$ between the two systems $A$ and $B$ with various sizes. Their behavior depends on the difference of sites $\Delta M =M_A-M_B=0,2,6,30, M_B+2$.  The system $B$ has a fixed number of sites, $M_B=100$, the temperatures and chemical potentials of both systems are $\beta_A=\beta_B=\beta=10 /\gamma$ and $\mu_A= \mu_B = 0.2 \gamma$, respectively. The coupling strength between $A$ and $B$ is $\gamma_C = 0.1 \gamma$. The time is scaled by the round trip time $\tau_B = M_B \hbar/\gamma $. Heat is given in units of thermal energy $\beta^{-1}$ and the number of exchanged particles is unscaled. }
	\label{fig1}
\end{figure}

Figure~\ref{fig1} illustrates the spontaneous transport of average heat (left panel) and average exchanged particle numbers (right panel) for equal temperatures ($\beta_A = \beta _B = \beta$) and equal chemical potentials $\mu_A = \mu_B= \mu$ as a 
function of time. Here, the heat is given in units of thermal energy, $\beta^{-1}$, while the number of exchanged particles
is unscaled. The time is given in units of $\tau_B = M_B \hbar/\gamma $ which is the shortest round trip time of a particle moving at maximum group velocity in the system $B$~\cite{euijin2016}. Note, that the part $B$ is assumed to be smaller than $A$, and  hence round-trip time of $B$ is shorter than that of $A$.      

Between the systems A and B with the same number of sites, neither heat nor particles can flow (see the red line in Fig.1). 
Because the two parts of the system as well as their initial states are identical, no preferred direction of flow exists.
Even if $M_A$ and $M_B$ differ from each other, heat or particles do not flow before $\tau=\tau_B$. This is so because for times less than the round trip time $\tau_B$, the system can be considered as infinite and symmetric, 
as discussed in the previous work \cite{euijin2016}.
Only for $\tau>\tau_B$, the finiteness of the systems manifest itself, and the asymmetry becomes apparent. 

\begin{figure}[tbp]
	\centering
	\includegraphics[width=\linewidth]{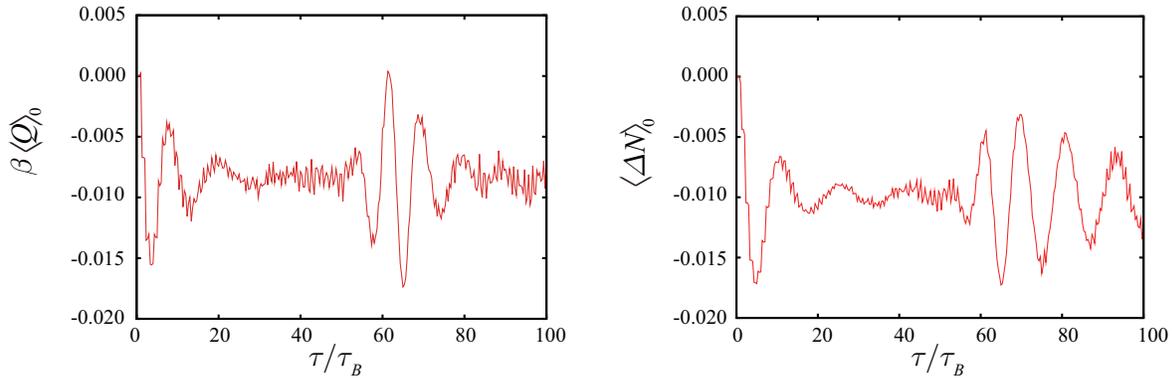}
	\caption{The large time behavior of the spontaneous heat (left panel) and particle (right panel) flow at vanishing affinity biases for the parameter values $\Delta M=2$, $\beta=10 \gamma$. }
	\label{fig2}
\end{figure}

The green and blue lines in Fig. \ref{fig1} display particle and heat transport when the systems differ in size by two and six, respectively.
For times larger than $\tau_B$, almost instantly a non-vanishing amount of spontaneously transmitted heat and particle number exchange sets in. Both heat and particle number remain almost constant up to the time $2 \tau_B$ when they again change in an almost step-like manner. At larger times a partial reversal and more erratic behavior of the transferred heat and particle number can be observed as illustrated in Fig.~\ref{fig2} for $M_A-M_B=2$. However, also for larger contact times heat and particles are always transferred from the larger to the smaller system. 
 
For larger size differences, such as $\Delta M=M_A-M_B=30$, the transferred heat and particle number exchange becomes considerably smaller (violet lines in Fig.1).
It can, however, recover larger values even for large $\Delta M$ if $M_A$ and $M_B$ satisfy the commensurability 
condition $p M_B = q M_A+O(1)$ with integers $p,q$.
For example, if $M_A=2M_B+2$  (cyan line in Fig.\ref{fig1}), $\Delta M=M_B+2$ is large, but $M_A$ and $M_B$ satisfy the commensurability condition with $p=2$ and $q=1$, a relatively large amount of particles and heat flows spontaneously.
In this case, the system behaves similarly as for $\Delta M=2$, 
apart from the latent period, which is twice as long, before the spontaneous transport sets in.
The similarity of the $\Delta M=2$ and $M_A=2M_B+2$ cases, can also be seen in the transmission property, which has previously been studied~\cite{euijin2016}.

In the presence of an affinity bias, spontaneous and bias-induced heat and matter transport contribute additively in agreement with Eqs. (\ref{Q1}) and (\ref{N1}). This leads to an increase of heat in proportion to the time $\tau$ up to $\tau_B$ where the slope suddenly changes to remain constant up to $2 \tau_B$, see Fig.~3 of Ref. \cite{euijin2016}. 
\begin{figure}[tbp]
\centering
\includegraphics[width=\linewidth]{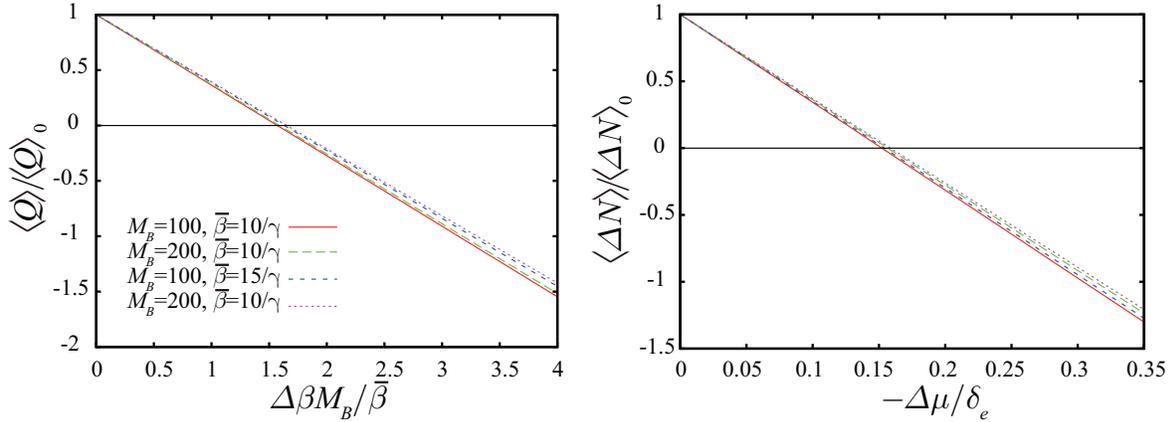}
\caption{
The heat $\langle Q \rangle$ and particle number $\langle \Delta N \rangle$ that are exchanged between systems of sizes $M_B = 100, 200$ and $\Delta M =2$ during the contact time $\tau = 1.5 \tau_b$  are scaled by the according instantaneous values  $\langle Q \rangle_0$ and  $\langle \Delta N \rangle_0$. 
These ratios are displayed in panel (a) and (b), respectively, as functions of the corresponding scaled affinity biases $\Delta \beta M_B/ \bar{\beta}$ and $- \Delta \mu / \delta_e$ where $\delta_e = 2 \gamma/M_B$ denotes the level spacing of $H_B$ near the band center. 
The average temperature is chosen as $\bar{\beta} = 10/\gamma,\; 15/\gamma$; the coupling strength is $\gamma_C=0.1\gamma$.
At smaller affinity biases the spontaneous contribution dominates the direction of transport pointing in the  opposite direction dictated by the corresponding biases. The dependence on the size and the average temperature is rather weak. 
}
\label{spscale}
\end{figure}

The contribution of the spontaneous heat flow remains significant in comparison to the affinity biased contributions as long as $\Delta \beta$ is sufficiently small. 
Figure~\ref{spscale} displays the total amount of exchanged heat  in panel (a) and particle number in panel (b) relative to the respective spontaneous values for systems with $\Delta M =2$ and for a contact time $\tau = 1.5 \tau_B$. While the spontaneously transmitted heat is transferred from the larger A to the smaller B part, a positive inverse temperature bias favors the transfer in the opposite direction. For relative inverse temperature differences $\Delta \beta/\overline{\beta} \lessapprox 1.5/M_B$ the spontaneous contribution dominates in determining the direction. Only at larger temperature differences the heat flows in the expected direction from hot to cold. A similar behavior can be observed for the transferred particle number which is oppositely oriented to the ``common'' direction as long as $\Delta \mu \gtrapprox -1.5 \delta_e$ where $\delta_e = 2  \gamma/M_B$ is the level-spacing near the band center. Both the relative transferred heat $\langle Q \rangle/\langle Q \rangle_{0}$ and particle number $\langle \Delta N \rangle/\langle \Delta N \rangle_0$ as functions of $\Delta \beta M_{B}/\bar{\beta}$ and of the scaled chemical potential difference $\Delta \mu / \delta_e$ only insignificantly depend on the temperature and the size of the part B. 
\begin{figure}[tbp]
\centering
\includegraphics[width=\linewidth]{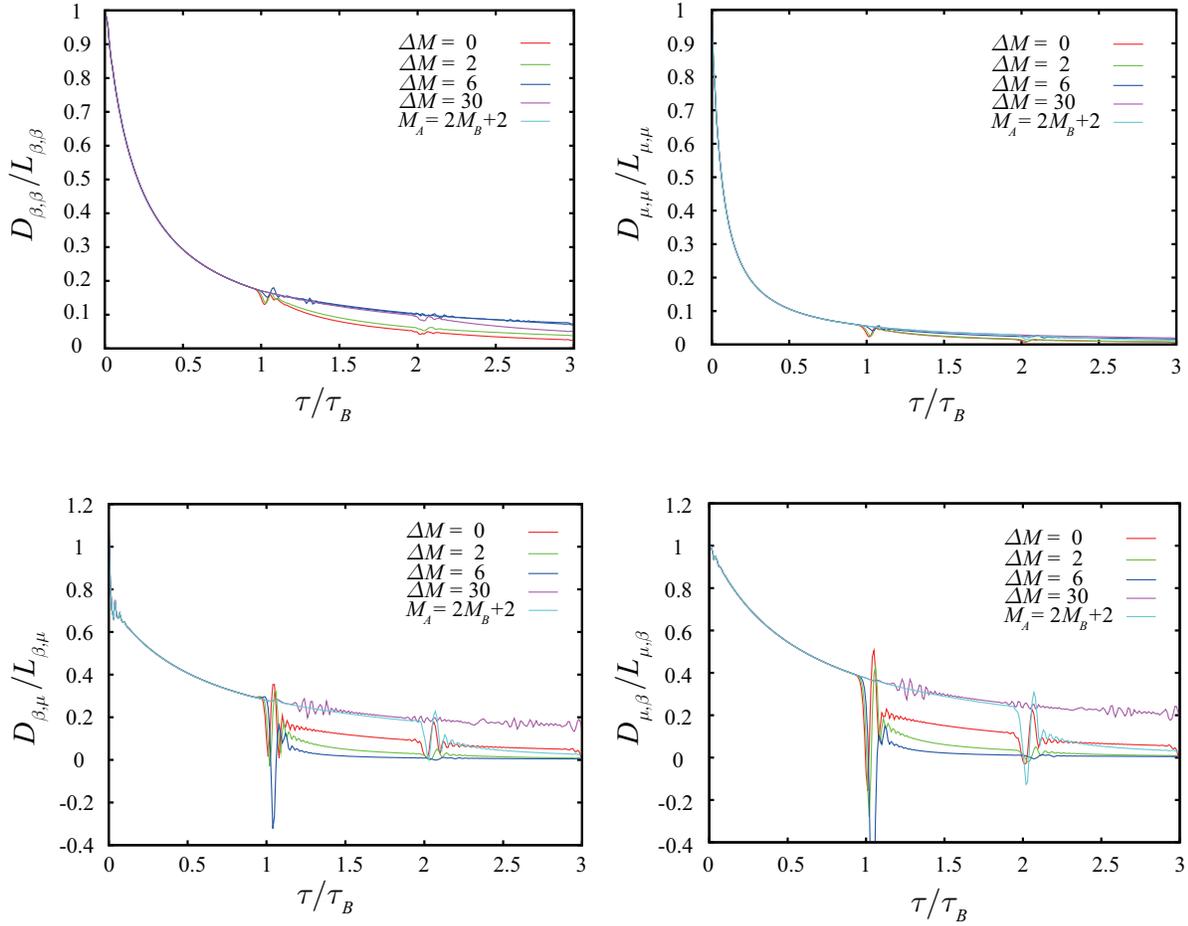}
\caption{The deviation $D=L-C$ relative to the positive and symmetric reference matrix $L$ is component-wise displayed as a function of the scaled time $\tau/\tau_B$ for $\Delta M =0,2,6, 30 $ and $M_A=2 M_B  +2 $. The other parameters are the same as in Fig. \ref{fig1}.}
\label{NewFig4}
\end{figure}

\begin{figure}[htp]
\centering
\includegraphics[width=\linewidth]{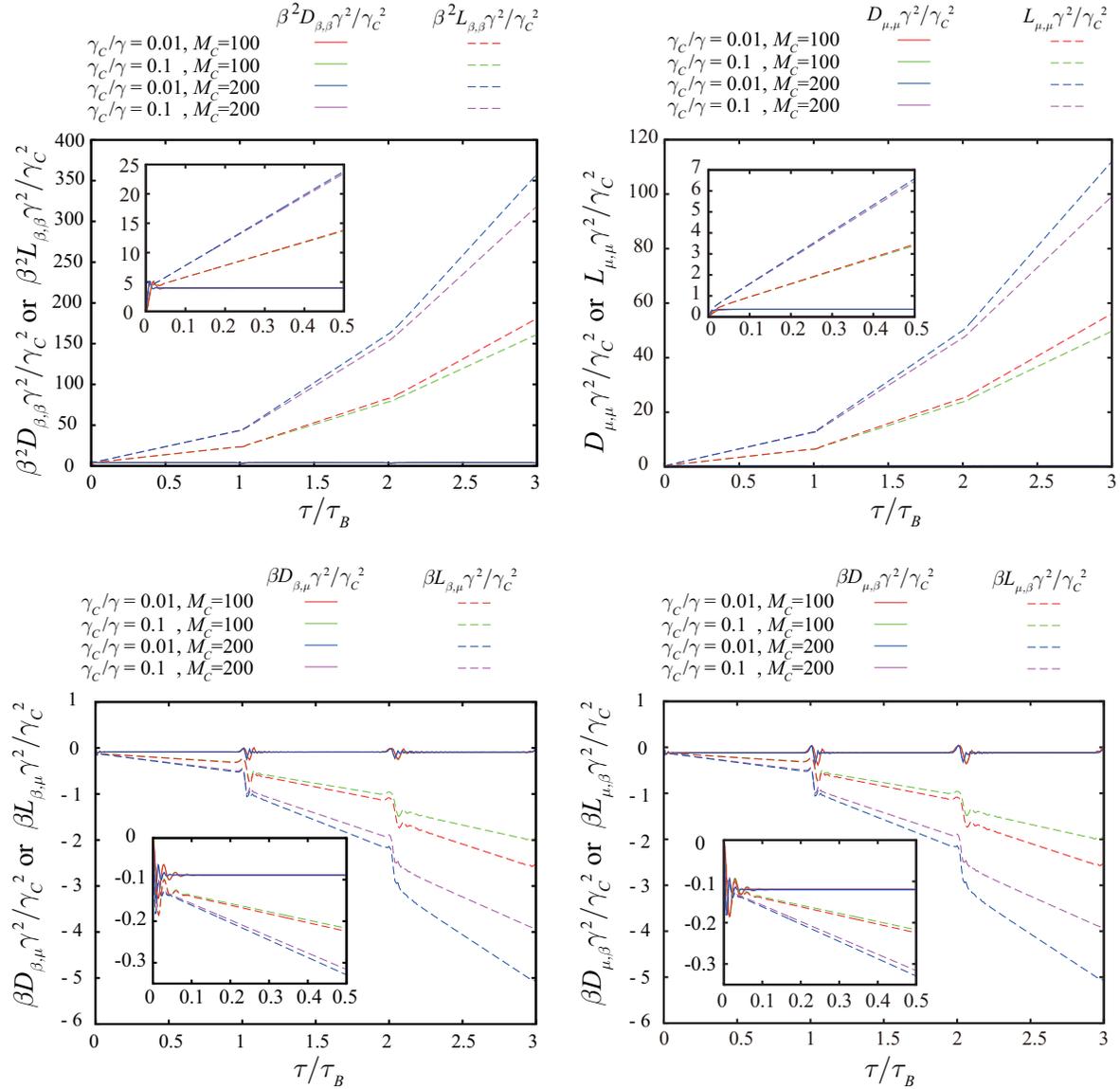}
\caption{The components of the matrices $D$ and $L$ 
are displayed as functions of the scaled contact time $\tau/\tau_B$ for different interaction strengths  $\gamma_C=0.01\gamma,0.1 \gamma$, and system sizes $M_A=M_B=100,200$ at the inverse temperature ${\beta}=10/\gamma$. The diagonal $(\beta,\beta)$- and  $(\mu,\mu)$-components 
are given in units of $\gamma_C^2/\gamma^2\beta^2$ and $\gamma_C^2/\gamma^2$ respectively, and the non-diagonal 
$(\beta,\mu)$- and $(\mu,\beta)$-components in units of $\gamma_C^2/\gamma^2\beta$. The insets present the respective behavior for short contact times 
$0<\tau<0.5\tau_B$.  }
\label{NewFig5}
\end{figure}

\begin{figure}[htp]
\centering
\includegraphics[width=\linewidth]{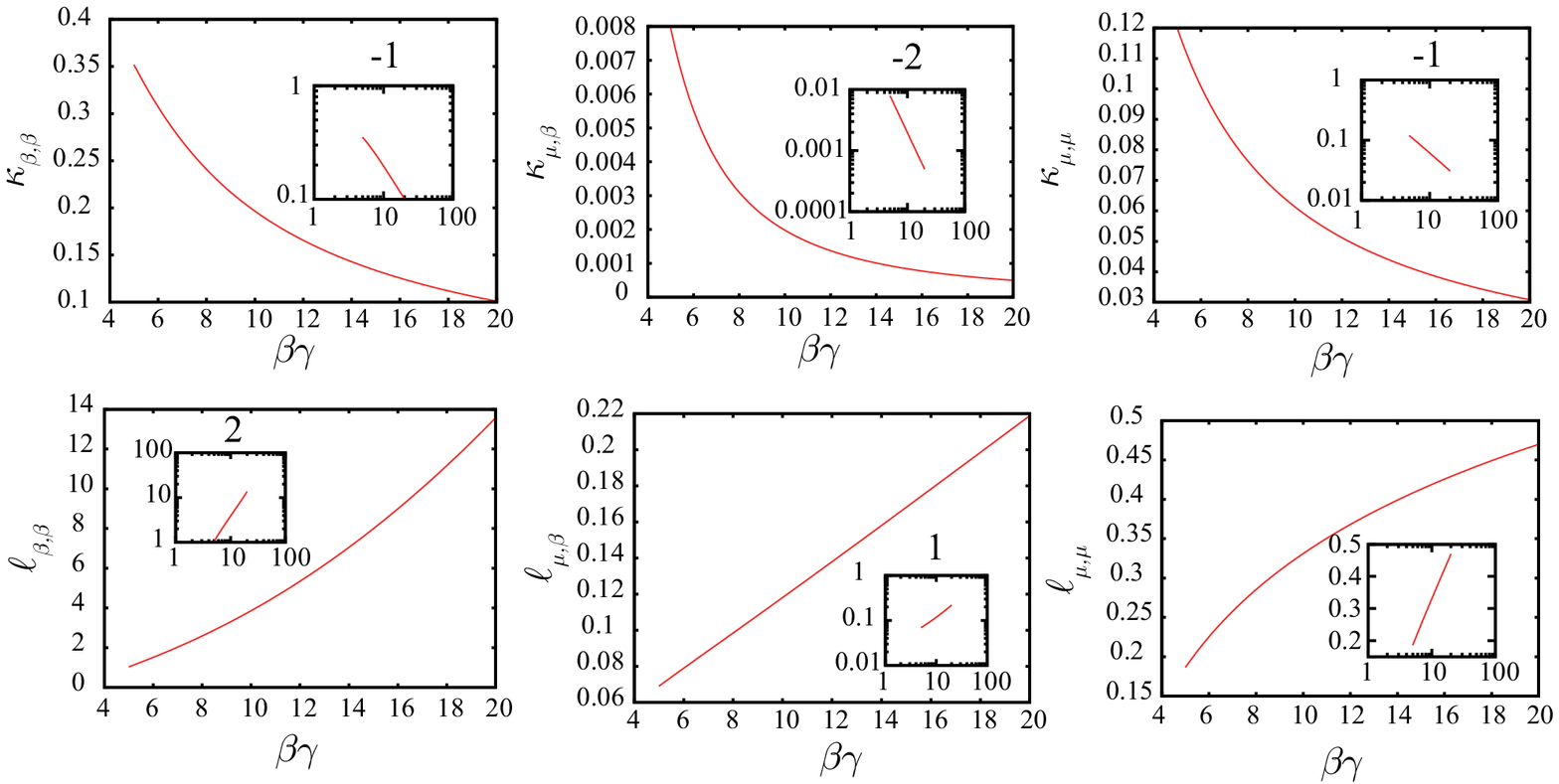}
\caption{ The dimensionless coefficients
$\kappa_{\chi,\eta}$ and $\ell_{\chi,\eta}$ specifying the matrix $L$ for times $\tau < \tau_B$ according to the Eqs. (\ref{Lbb} -- \ref{Lmm}) are represented as
functions of $\beta \gamma$. In all cases but for $\ell_{\mu,\mu}$ the coefficients depend algebraically on $\beta \gamma$ as can be seen from the doubly logarithmic plots in the insets yielding
$\kappa_{\beta,\beta}\propto (\beta\gamma)^{-1}$,  $\kappa_{\beta,\mu}\propto (\beta\gamma)^{-2}$,  $\kappa_{\mu,\mu}\propto (\beta\gamma)^{-1}$,  $\ell_{\beta,\beta}\propto (\beta\gamma)^2$,  $\ell_{\beta,\mu}\propto \beta\gamma$.
The remaining coefficient is well approximated as $\ell_{\mu,\mu}\propto \log (\beta\gamma)$. The numbers above the insets indicate the respective exponent that is determined from the slope of the doubly logarithmic graph displayed in the inset. 
}
\label{fig6}
\end{figure}

The impact of the work on the bias-induced transport properties can be quantified by the matrix $D$ introduced in Eq. (\ref{D}) which is  the difference between the actual transport matrix $C$, see Eq. (\ref{C}), and the symmetric and positive reference matrix $L$, defined in Eq. (\ref{LXX}), i.e. $D=L-C$.  
Figure~\ref{NewFig4} displays the elements of the deviation matrix $D$ normalized by the respective elements of the matrix $L$ as a functions of the contact time $\tau$.

Starting from an initially large value, the relative deviation decays as the contact time increases. 
As already mentioned,  for contact times less than $\tau_B$ the finiteness of the system is not effective due to the finite propagation speed of the perturbation caused by the contact. Hence the decay behavior  is independent of the size $M_B$ and the difference $\Delta M$ up to $\tau_B$. Only at larger contact times, a weak dependence of the decay of the matrix elements $D_{\chi,\eta}$ on $M_B$ and $\Delta M$ can be observed, where both $\chi$ and $\eta$ may denote $\beta$ or $\mu$.
The decay of the relative deviations $D_{\chi,\eta}/L_{\chi,\eta}$ is solely caused by the growth behavior of the matrix elements of $L$ while the difference matrix $D$ is essentially constant apart from a very short interval near $\tau=0$ as shown in Fig. \ref{NewFig5}.
Within this short interval of small contact times, the matrix elements of $D$ and $L$ perform rapid oscillations resulting in a small offset followed up by constant, $\tau$-independent values in the case of the matrix $D$, and by a linear increase of the diagonal elements of $L$ and a linear decrease of the off-diagonal element $L_{\beta,\mu}$ up to the contact time $\tau_B$. Within a small vicinity of $\tau_B$ the matrix elements of $D$ perform small oscillations to assume the previous value also for larger values of $\tau$. The diagonal elements of $L$ resume a constant growth rate, which, however, is larger for $\tau >\tau_B$ than it was for $\tau < \tau_B$. The off-diagonal element $L_{\beta.\mu}$ displays a rapid deflection near $\tau_B$ followed by a steeper decrease than before.
At the considered small values of the interaction parameter $\gamma_C$ the matrices $D$ and $L$ are proportional to $\gamma^2_C$ up to the contact time $\tau_B$. 
Therefore, for contact times $\tau < \tau_B$ the elements of the matrix $L$ are well approximated by the following linear laws: 
\begin{eqnarray}\label{Lbb}
L_{\beta,\beta}&=&(\gamma_C/\gamma)^2{\beta}^{-2}[\ell_{\beta,\beta}({\beta}\gamma)+\kappa_{\beta,\beta}({\beta}\gamma)\tau/\tau_0]\\
L_{\beta,\mu}&=&(\gamma_C/\gamma)^2{\beta}^{-1}[\ell_{\beta,\mu}({\beta}\gamma)+\kappa_{\beta,\mu}({\beta}\gamma)\tau/\tau_0] \label{Lbm}\\
L_{\mu,\mu}&=&(\gamma_C/\gamma)^2[\ell_{\mu,\mu}({\beta}\gamma)+\kappa_{\mu,\mu}({\beta}\gamma)\tau/\tau_0] \label{Lmm}
\end{eqnarray}
with $\tau_0=\hbar/\gamma$ being the hopping time between neighboring sites. This time is independent of the size of the system.
The coefficients $\ell_{\chi,\eta}$ and 
$\kappa_{\chi,\eta}$ are dimensionless functions of ${\beta}\gamma$, which are plotted in Fig.~\ref{fig6} for a range of low temperatures, ${\beta} \gamma >4$, which yet are large compared to the level-spacing $\delta_e$ such that ${\beta} \delta_e \ll 1$. 
Apart from $\ell_{\mu,\mu}$ which grows logarithmically as $\ell_{\mu,\mu} = 0.20 \ln (\beta \gamma)$ all other coefficients can be modeled in this temperature range by algebraic functions: $\ell_{\beta,\beta} =0.034 ({\beta} \gamma)^2$, $\ell_{\mu,\beta} = 0.011 {\beta} \gamma$, $\kappa_{\beta,\beta} = 2.0 ({\beta} \gamma)^{-1}$, $\kappa_{\mu,\beta}= 0.20({\beta}\gamma)^{-2}$ and $\kappa_{\mu,\mu} = 0.62({\beta} \gamma)^{-1}$.

\begin{figure}[tbp]
	\centering
	\includegraphics[width=\linewidth]{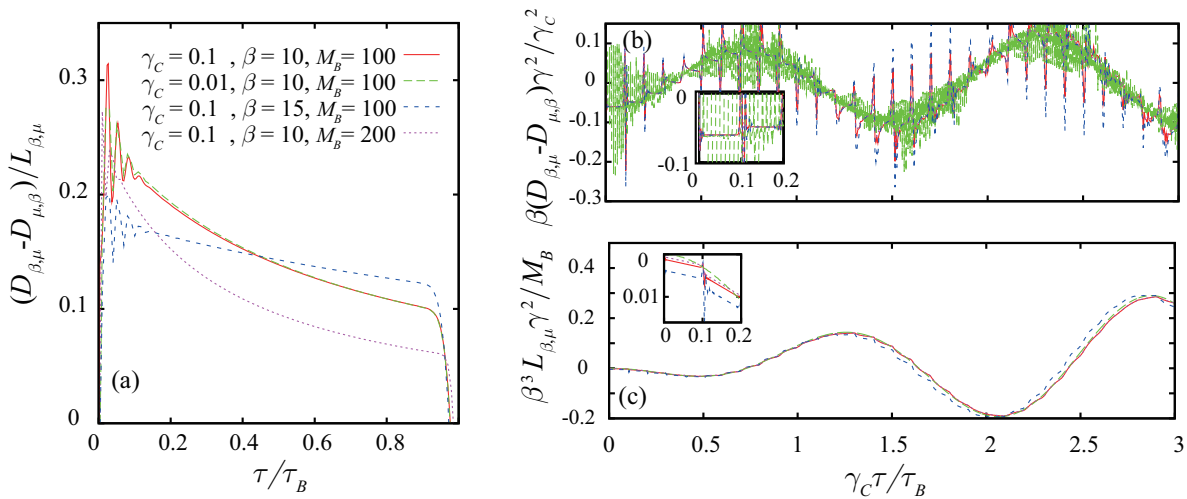}
	\caption{The  asymmetry of the transport coefficients is illustrated for symmetric systems ($\Delta M =0$) with various parameters $\gamma_C$, $\bar{\beta}$ and $M_B$ as specified in panel (a). In panel (a) the ratio of the Onsager asymmetry, $D_{\mu,\beta}-D_{\beta,\mu}$, and the reference, symmetric transport coefficient $L_{\mu,\beta}$, is displayed as a function of time $\tau/\tau_B$ in range of $0<\tau<\tau_B$. 
In panel (b) the asymmetry is scaled in units of $\beta^{-1} (\gamma_c/ \gamma)^2$ and the time in units of $\tau_B\gamma /\gamma_C$ yielding an approximate data collapse for different $\gamma_C$, $\beta$, and $M_B$ onto a periodic master function. 
The spikes are located at integer multiples of the  round trip time $\tau_B$. 
 The inset displays a magnification of the scaled asymmetry for times $0<\tau<0.2\tau_B/\gamma_C$.  The non-diagonal element $L_{\beta,\mu}$ 
is presented in units of $M_B/\beta^3\gamma^2$ as a function of the scaled time $\gamma_C\tau/(\gamma \tau_B)$ as evidenced in panel (c). 
The inset exhibits the behavior at short times. 
}

	\label{fig4}
\end{figure}

 The $\tau$-independent contributions of $\ell_{\beta,\beta}$ and $\ell_{\beta,\mu}$ are dominant for very small contact times, $\tau \ll \tau_B$, rendering the matrix elements $L_{\beta,\beta}$ and $L_{\beta,\mu}$ almost independent of ${\beta}$. 
In this regime of very small $\tau$, the matrix elements $D_{\chi,\eta}$ and $L_{\chi,\eta}$ have similar values, 
and therefore the ratios $D_{\chi,\eta}/L_{\chi,\eta}$ are close to 1 as illustrated in Fig.~\ref{NewFig4} for ${\beta} \gamma =10$. The same behavior can also be observed for other temperature values with ${\beta} \gamma >4$. 
According to Fig.~\ref{NewFig5}, in the contact time window $0 < \tau <\tau_B$,  the behavior of $D$ can well be represented as 
$D_{\beta,\beta}\simeq 0.034\gamma_C^2$, $(D_{\mu,\beta} +D_{\beta,\mu})/2\simeq 0.011\gamma_C^2/\gamma$ and $D_{\mu,\mu}\simeq0.20\ln({\beta}\gamma)$. For a discussion of the  asymmetry of $D$ we refer to Fig.~\ref{fig4} and to the according text below it. 
For sufficiently large values of  $\tau$,  which are still less than $\tau_B$, the $\tau$-independent contributions to $L$ can be neglected yielding 
$L_{\beta,\beta}\simeq 2 (\gamma_C/\gamma)^2{\beta}^{-3}\gamma^{-1}\tau/\tau_0$, $L_{\beta,\mu}\simeq 0.2 (\gamma_C/\gamma)^2{\beta}^{-3} \gamma^{-2} \tau/\tau_0$, 
and $L_{\mu,\mu}\simeq 0.62 (\gamma_C/\gamma)^2 ({\beta} \gamma)^{-1}  \tau/\tau_0$.

For the same range of contact times $\tau < \tau_B$ excluding very short ones,  
the diagonal elements and the symmetrized non-diagonal element of   
the actual transport coefficients follow from 
$C_{\chi,\eta}=L_{\chi,\eta}-D_{\chi,\eta}$ with the above expressions for the matrix elements of $L$ and $D$. 
For contact times $\tau > \tau_B$ the dependence  of $L$ on $\tau$ becomes non-linear. Though, a  scaling of $L(\tau/\tau_B) \propto M_B$ continues to hold.
In this non-linear region, small deviations from the proportionality to $\gamma_C^2$ can be observed for the matrix $L$ while $D$ remains unaffected.

Finally we focus on the violation of the reciprocity relation which can be quantified in terms of the difference of the non-diagonal elements of the matrix $D$, which, due to the symmetry of the matrix $L$, coincides with the respective difference of the non-diagonal elements of $C$. We consider symmetric systems with equal numbers of sites, $M\equiv M_A=M_B$. In panel (a) of Fig.~\ref{fig4} the relative degree of asymmetry,  $(D_{\mu,\beta}-D_{\beta, \mu})/L_{\beta,\mu}$, is displayed as function of $\tau/\tau_B$ for different values of the coupling strength $\gamma_C$, inverse temperature $\beta$, and size $M_B$. 
For short contact times $\tau$ it shows an oscillatory behavior which turns into a monotonic decay. 

While, at the considered low temperatures $\beta \gamma>4$,  the diagonal ($D_{\beta,\beta}$, $D_{\mu,\mu}$) and the symmetrized $(D_{\beta,\mu}+D_{\mu,\beta})/2$ 
 of $D$ are virtually independent of temperature, the difference of the non-diagonal elements $D_{\beta,\mu}-D_{\mu,\beta}$ grows proportional to the temperature as can be inferred from the data collapse displayed in Fig.~\ref{fig4}(b) upon a scaling of the difference by $ (\gamma_c/\gamma) \beta^{-1}$.  
The decrease of the relative difference $(D_{\beta.\mu}-D_{\mu,\beta})/L_{\beta,\mu}$ becomes faster with increasing system size $M_B$ due to the proportionality of $L$ to $M_B$ and the $M_B$-independence of $D$.    
 The dependence of the relative asymmetry degree on the coupling constant $\gamma_C$ is rather insignificant on short times because both matrices $D$ and $L$ are proportional to $\gamma_C^2$ in leading order. 

For larger contact times $\tau>\tau_B$, higher orders of the coupling strengths contribute to the matrices $D$ and $L$. Yet by using $\tau_C\equiv \tau_B \gamma/\gamma_C$ as a unit of time and, as mentioned above, scaling the difference $D_{\beta,\mu}- D_{\mu,\beta}$ with $(\gamma_C/\gamma)^2\beta^{-1}$ one finds a collapse of the off-diagonal elements onto a single curve, apart form a superposition of spikes at integer multiples of $\tau_B$ as displayed in panel (b) of Fig.~\ref{fig4}. Likewise an almost perfect data collapse is found in panel (c) for the off-diagonal element $L_{\beta,\mu}$ scaled by $M_B/(\beta^3 \gamma^2)$ as a function of  $\tau/\tau_C$. The displayed large time behavior, however, will depend on the specific nature of the interacting parts $A$ and $B$. For example for systems that equilibrate after a sufficiently large time the matrices $C$ and $L$ are expected to approach values independent of the contact time.
In any case, at times larger than $\tau_B$ the matrices $C$ and $L$ are no longer linearly proportional to the contact time. Therefore a comparison with the transport behavior following from standard transport theory is no longer meaningful.

We conclude that, in  agreement with our general analysis, the transport of heat and particle numbers between two linear systems described by the tight binding Hamiltonians (\ref{Ha}) deviates from the standard behavior by the presence of spontaneous transport occurring in the absence of any affinity bias, and by an asymmetry of the transport matrix signaling a violation of Onsager's reciprocity relation.  

\section{Summary}
We scrutinized the basic assumptions underlying the Onsager relations for the transport of heat and particles in relatively small systems. Our analysis is based on the standard assumption, see e.g. \cite{andrieux2009}, that the two parts of a system are prepared in grand-canonical equilibrium states with generally different temperatures and chemical potentials. In this initial state the energies and the particle numbers are separately determined, and then the two parts are brought into contact such that energy and particles can be exchanged between them. After a prescribed time $\tau$ the interaction of the two parts is switched off and energies and particle numbers are measured again. 

In contradistinction to standard transport theory \cite{deGrootMazur} we found spontaneous transport in the absence of a temperature and chemical potential difference of the two systems and also deviations from the reciprocity relations. Both effects have their origin in the fact that with bringing the parts of the system into contact and separating them again, work is done on the total system. Only if this work identically vanishes these deviations exactly  disappear and Onsager's standard transport theory follows from a fluctuation theorem. 
The spontaneous transport becomes visible only after a characteristic time which grows with the size of the system. The deviation from the reciprocity relations however, is most pronounced during this initial period.   
The presence of work is tantamount to the breaking of time-translational symmetry. This leads to non-symmetric transport coefficients and hence a violation of Onsager's reciprocity relations.
The same conclusions also hold for the alternative definitions of affinity biases and corresponding transport quantities as described in Ref.~\cite{alternative}. 

Both the general analysis presented in the first part of this work as well as the illustrative example are expressed in quantum mechanical terms. 
However, also systems governed by the laws of classical mechanics 
experience a change of energy imposed by contacting and disconnecting the parts of the system. Therefore we expect that the deviations of the transport properties from standard transport theory also apply for small classical systems.  

{\it Acknowledgments.} This work was supported by the National Research Foundation of Korea (NRF) grant funded by the Korea government (MSIP) (Grant No. NRF-2017R1A2B4007608)
and by the Deutsche Forschungsgemeinschaft via the projects HA 1517/35-1 and DE 1889/1-1.

%\end{linenumbers}


\begin{thebibliography}{10}
\bibitem{alternative} With the alternative definition $\Delta  \beta \mu = \beta_A \mu_A - \beta_B \mu_B$ for the chemical affinity bias the corresponding expression for the heat becomes $\tilde{Q} = \delta E_A - \Delta E_B$. The exponent on the right hand side of Eq. (\ref{PQ}) then assumes the form  $\bar{\beta} W +\tilde{Q} \Delta \beta - \Delta N \Delta \beta \mu$.
\bibitem{timereversal} For the sake of simplicity we assume that $\theta \hat{H} \theta = \hat{H}$, where $\theta$ is the time reversal operator. Hence we exclude the presence of fields that transform oddly under time-reversal such as magnetic fields and Coriolis forces.
\end{thebibliography}
\end{document}